\documentclass[aps,prd,twocolumn,groupedaddress,10pt]{revtex4-1}
\usepackage{amsmath, amsfonts, graphicx,placeins}
\usepackage{placeins}

\begin{document}

\title{Evaluation of low-$Q^2$ fits to $ep$ and $ed$ elastic scattering data}

\author{Timothy B. Hayward}
\author{Keith A. Griffioen}
\affiliation{Physics Department, College of William and Mary, Williamsburg, Virginia 23187}

\date{\today}

\begin{abstract}
We examined several low-$Q^2$ elastic $ep$ and $ed$ scattering data sets using various models to extract the proton and deuteron rms radii and developed a comprehensive algorithm for estimating the systematic bias of each extracted radius. In each case, we chose the model and upper bound for $Q^2$ that minimized the combination of statistical uncertainty and bias. We attribute the discrepancy between small ($\approx 0.84$~fm) and large ($\approx 0.88$~fm) proton radius extractions to the absence of data that can accurately isolate the linear and quadratic contributions to the form factor at low-$Q^2$. In light of this ambiguity, we estimated the asymmetric model-dependence of each extracted radius by studying the distribution of many possible fits. The resulting radii are 0.842(4)~fm for the proton and 2.092(19)~fm for the deuteron. These values are consistent with those determined from muonic hydrogen measurements within 0.3$\sigma$ for the proton and 1.8$\sigma$ for the deuteron.
\end{abstract}



\maketitle

\section{Introduction}
The rms radius of the proton has been measured by three different techniques: $ep$ elastic scattering at low momentum transfer $Q^2$ \cite{Hand1963, Yerevan1972, Saskatoon1974, Mainz1974, Mainz1975,Mainz1980, Mainz2010},  hydrogen Lamb shifts \cite{hyd_lamb1, hyd_lamb2, hyd_lamb3, hyd_lamb4, hyd_lamb5} and muonic hydrogen Lamb shifts \cite{Pohl2010,Antognini2013}. The CODATA average combines the first two methods, yielding a radius of 0.875(6)~fm~\cite{CODATA}. This is more than 5$\sigma$ larger than the muonic Lamb shift value of 0.84087(39)~fm.

In elastic scattering, the extracted radius depends on the slope of the electric form factor at zero momentum transfer. Experiments, however, can only measure the shape of the form factor at finite values. Hence, the extracted radius depends on the reliability of an extrapolation and the assumptions made in fitting. With hindsight of the muonic hydrogen results, several reanalyses of existing $ep$ data~\cite{smallRadii00,smallRadii1,smallRadii2,smallRadii3,smallRadii4,smallRadii5} have extracted ``small'' proton radii, consistent with $R_E \approx 0.84$~fm. Using the same data, the reanalyses of Refs. \cite{largeRadii00, largeRadii1,largeRadii2,largeRadii3,largeRadii4,largeRadii5} yield ``large'' radii values consistent with CODATA.

The deuteron radius has been measured using $ed$ elastic scattering~\cite{Simon1980,Platchkov1990} and muonic deuterium Lamb shifts~\cite{Pohl:2016glp}. In this case the CODATA value of 2.142(2)~fm differs from the muonic Lamb shift result of 2.1256(8)~fm by more than 8$\sigma$.

With these discrepancies in mind, we have developed a fitting algorithm to decide the best fit function and upper bound in $Q^2$ that minimizes bias (systematic shifts in the extracted radius from too rigid fit models) and variance (statistical uncertainty of the extracted radius). We applied this algorithm to several extant data sets. 

\section{Formalism}
\subsection{The Proton}

The 4-momentum transfer squared for an electron scattering from an atomic nucleus at rest is given by
\begin{align}
Q^2 &= -q^2 = 4EE'\sin^2\frac{\theta}{2},
\end{align}
in which $E$ is the initial electron energy, $E'$ is the outgoing electron energy and $\theta$ is the electron scattering angle. 

In the Born approximation, the elastic $ep$ scattering cross section can be written in terms of the Sachs electric and magnetic form factors, $G_E(Q^2)$ and $G_M(Q^2)$, as
\begin{align}
\frac{d\sigma}{d\Omega} &= \left( \frac{d\sigma}{d\Omega} \right)_{\text{Mott}}\frac{1}{1+\tau} \left[ G_E^2(Q^2) + \frac{\tau}{\epsilon} G_M^2(Q^2)\right].
\end{align}
The Mott cross section is given by
\begin{align}
 \left( \frac{d\sigma}{d\Omega} \right)_{\text{Mott}} &= \frac{4 \alpha^2 \cos^2\frac{\theta}{2} {E'}^3}{Q^4 E},
\end{align}
in which $\alpha$ is the fine-structure constant,
\begin{align}
\tau &= \frac{Q^2}{4M^2},
\end{align}
$M$ is the proton mass and
\begin{align}
\epsilon &= \left( 1+2(1+\tau)\tan^2 \frac{\theta}{2}\right)^{-1}.
\end{align}
The form factors are normalized at $Q^2~=~0$ such that
\begin{align}
G_E(0)&=1
\end{align}
and
\begin{align}
G_M(0) &= \mu_p \approx 2.793,
\end{align}
the proton's magnetic moment.

Determining the proton charge radius requires extrapolating the electric form factor to $Q^2=0$ and determining the slope at the origin. The rms radius, $R_E$, is given by the second term in the low-$Q^2$ expansion,
\begin{align}
G_E(Q^2) &= 1- \frac{1}{6}R_E^2Q^2+c_2 Q^4 + ...
\end{align}
Therefore,
\begin{align}
R_E^2 &= -6 \frac{dG_E}{dQ^2} \bigg|_{Q^2\rightarrow0}.
\end{align}

\subsection{The Deuteron}

The elastic $ed$ scattering cross section can be written as
\begin{align}
\frac{d\sigma}{d\Omega} &= \left( \frac{d\sigma}{d\Omega} \right)_\text{Mott} \left[ A(Q^2) + B(Q^2) \tan^2 \frac{\theta}{2} \right].
\end{align}
The two structure functions, $A(Q^2)$ and $B(Q^2)$, are combinations of the charge, $G_C(Q^2)$, magnetic, $G_M(Q^2)$, and quadrupole, $G_Q(Q^2)$, form factors such that
\begin{align}
A(Q^2) &= G_C^2(Q^2) + \frac{2}{3} \eta G_M^2(Q^2) + \frac{8}{9} \eta^2 G_Q^2(Q^2)
\end{align}
and
\begin{align}
B(Q^2) &= \frac{4}{3} \eta (1+\eta) G_M^2(Q^2),
\end{align}
in which
\begin{align}
\eta = \frac{Q^2}{4M_d^2}
\end{align}
and $M_d$ is the deuteron mass. The deuteron form factors are normalized at $Q^2 = 0$ in a similar manner to the proton:
\begin{align}
G_C(0) &= 1,\\
G_M(0) &= \frac{M_d}{M}{\mu_d} \approx 1.714
\end{align}
and 
\begin{align}
G_Q(0) &= M_d^2 Q_d \approx 25.83,
\end{align}
in which $M$ is still the proton mass, $\mu_d$ is the magnetic dipole moment \cite{Mohr} and $Q_d$ is the electric quadrupole moment \cite{Ericson}.
The deuteron charge radius is determined from the slope of the charge form factor at $Q^2=0$:
\begin{align}
R_E^2 &= -6 \frac{dG_C}{dQ^2} \bigg|_{Q^2\rightarrow0},
\end{align}
or equivalently, using the slope of $A(Q^2)$ at $Q^2=0$,
\begin{align}\label{deuteron_radius}
R_E^2 &= -3 \frac{dA}{dQ^2} \bigg|_{Q^2\rightarrow0} + \frac{G_M^2(0)}{2 M_d^2},
\end{align}
in which
\begin{align}
 \frac{G_M^2(0)}{2 M_d^2} \approx 0.0163~\text{fm}^2.
\end{align}
We used Eq. \ref{deuteron_radius} to extract the deuteron radius. The advantage of using $A(Q^2)$ for extracting $R_E$ is that no subtraction of $G_C$ and $G_M$ is required. As long as the fit is robust enough to capture the form of the structure function at the origin, the radius can be determined from $A(Q^2)$ alone.

\section{Models}\label{models}

\subsection{Normalization}\label{Normalization}

State-of-the-art form factor measurements can determine relative cross sections to $\approx 0.1\%$ but absolute cross sections only to $\approx 1\%$. Thus, all reasonable fits require an overall normalization constant. This was demonstrated in the Mainz (1975)~\cite{Mainz1975} analysis. These researchers fit their data to the form $c_0 + c_1 Q^2$ and extracted $c_0 = 0.994(2)$ and $R_E = 0.84(2)$~fm. They then demonstrated the effect of forcing $G_E(0)=1$, with no penalty in their fit, using the form $1 + c_1 Q^2$, and extracted a radius of $0.88(2)$~fm. 

To demonstrate the effect of the normalization we generated pseudodata from $Q^2 = 0.005$ to $0.05$~GeV$^2$ with perfect statistics using the dipole form factor,
\begin{equation}\label{dipole_form_factor}
G_D(Q^2) = \frac{c_0}{\left(1+Q^2/0.71~\text{GeV}^2\right)^2},
\end{equation}
for values of $c_0$ between 0.99 and 1.01. This range represented the uncertainty in the overall cross section that a typical experiment may experience. We fit the pseudodata to linear functions with two different forms of normalization: $1+c_1 Q^2$ and $c_0 (1+c_1 Q^2)$. 

\begin{figure}[h!]
\includegraphics[scale=0.41]{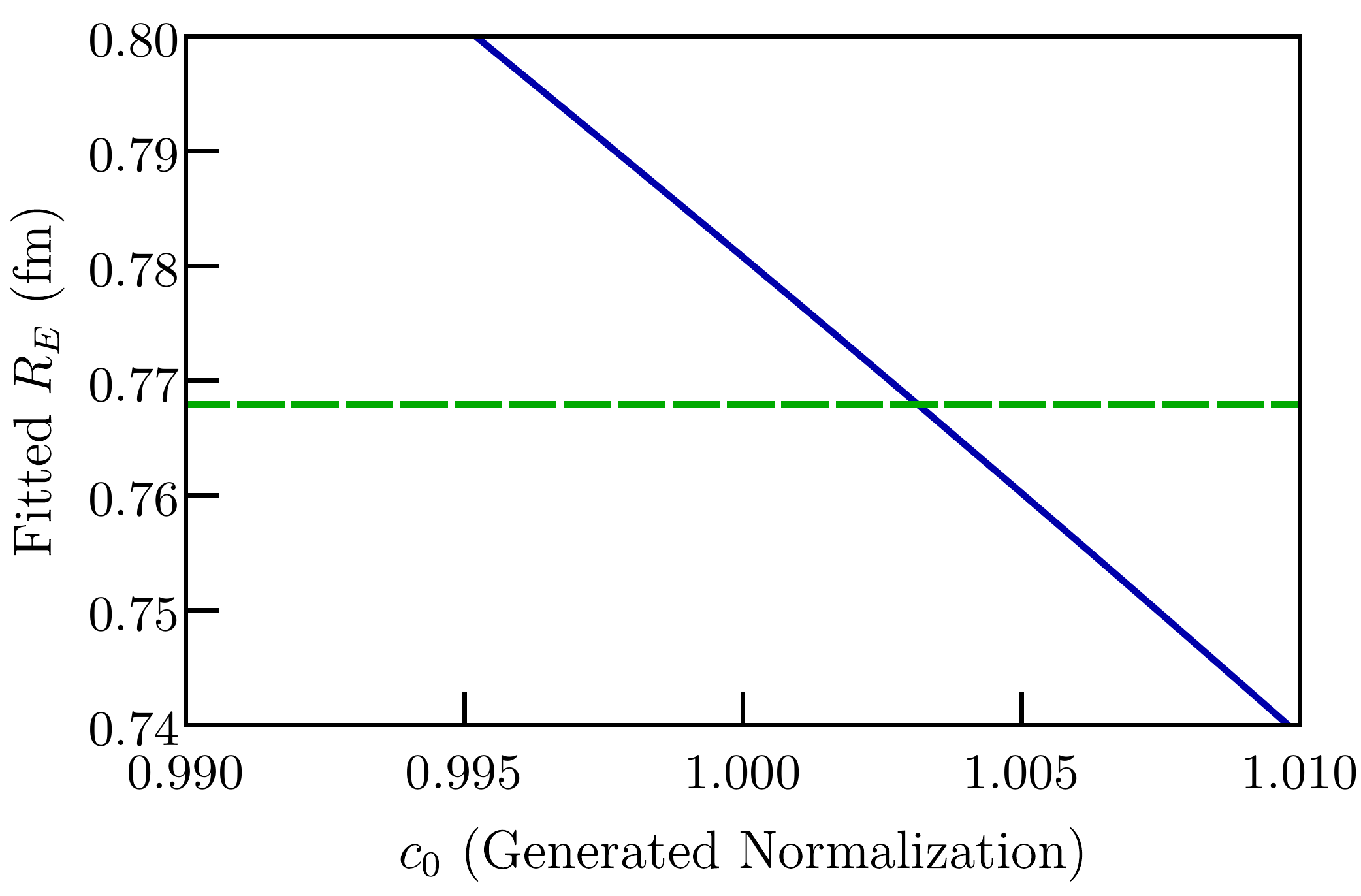}
\caption{(Color online). The extracted radius from $1+c_1Q^2$ (solid blue line) and $c_0(1+c_1 Q^2)$ (dashed green line) as a function of the generated normalization, $c_0$. The fit $c_0(1+c_1 Q^2)$ yields a stable result independent of the generated normalization. The dipole radius (0.811~fm) is not accurately reproduced because of bias in the fit form unrelated to the normalization.}
\label{normalization}
\end{figure}

FIG. \ref{normalization} shows the radius obtained from both cases as we changed the input value of $c_0$ in Eq. \ref{dipole_form_factor}. A fit function with an overall multiplicative normalization (green dashed line) yields a consistent radius independent of the experimental normalization. If the fit is forced through the point $Q^2(0)=1$ (solid blue line), the extracted radius can be biased by the uncertainty on the experimental normalization. The amount of bias will depend heavily on the range in $Q^2$ present in the data.

\subsection{Generating Functions}\label{generating_functions}
Correctly modeling the curvature of the form factor at low-$Q^2$ is crucial for obtaining an accurate proton radius. In order to study the effect that particular choices of fitting functions could have on the extracted proton radius we chose two benchmark models to simulate the form factor. These models encompass a range of curvatures expected in low-$Q^2$ form factor data and were used to provide estimates of the bias in our fits. To model the proton we used an exponential,
\begin{align}
{G_{E~\text{(exp)}}} (Q^2)&= \text{exp}\left({-\frac{R_E^2}{6}Q^2}\right),
\end{align}
which is known to fall off faster than the actual electric form factor, and a dipole,
\begin{align}
{G_{E~\text{(dip)}}} (Q^2)&= \left(1+\frac{R_E^2 Q^2}{12}\right)^{-2},
\end{align} 
which cannot reproduce the full curvature seen in the form factor but has often been used as a first approximation historically. For the deuteron, we used the Abbott~\cite{Abbott:2000ak} parameterization of the deuteron form factors as our model input. The Abbott radius, 2.094(3)~fm, is thought to be too small but the curvature can serve as a useful benchmark. 

\subsection{Fitting Functions}\label{fitting_functions}
Models with too few parameters cannot reproduce the shape of the data, but models with too many parameters can lead to fitting unphysical fluctuations. We attempted to incorporate a range of possible complexity in our fits and ultimately selected 11 functions. These functions were polynomials, 
\setlength\headheight{10pt}
\begin{align}\label{polynomials}
P_{n,0}(Q^2) = c_0 \left(1+ \textstyle{\sum_{i=1}^{n}} c_i Q^{2i} \right),
\end{align}
inverse polynomials,
\begin{align}\label{inversePolynomials}
P_{0,n}(Q^2) &= \frac{c_0}{1+\sum_{i=1}^{n} c_i Q^{2i}},
\end{align}
where $n = 1-4$, and three continued fraction functions,
\begin{align}\label{continuedFractions}
{CF}_2(Q^2) &=  \frac{c_0}{1+\frac{c_1 Q^2}{1+c_2 Q^2}},\\
{CF}_3(Q^2)&=\frac{c_0}{1+\frac{c_1 Q^2}{1+\frac{c_2 Q^2}{1+c_3 Q^2}}},\\
{CF}_4(Q^2)&=\frac{c_0}{1+\frac{c_1 Q^2}{1+\frac{c_2 Q^2}{1+\frac{c_3 Q^2}{1+c_4 Q^2}}}}.
\end{align}
We limited the complexity of our functions to quartic powers of $Q^2$ in order to keep the fit of $G_E$ well-behaved outside of the interpolating region \cite{smallRadii2}. 

\section{Fitting Procedure}\label{fitting_procedure}
\subsection{The Role of $\chi^2$}

Fits to experimental data typically rely on the minimization of
\begin{equation}\label{chi2function}
\chi^2 = \sum_i^N \left(\frac{G_E\left(Q_i^2 \right) - f(Q_i^2)}{\sigma_i} \right)^2,
\end{equation}
in which $G_E(Q_i^2)\pm \sigma_i$ are the data values and uncertainties, and $f(Q_i^2)$ are the parameterized fit values at $Q_i^2$. A good fit is expected to have $\chi^2/\text{dof}~\approx~1$, in which the degrees of freedom~(dof) are the number of data points minus the number of free parameters. But $\chi^2/\text{dof}$ can be misleading if experimental uncertainties are under- or over-estimated. Additionally, it is frequently possible to obtain different fits to the same data and still arrive at a satisfactory $\chi^2$. 

This can be demonstrated with the published Mainz (2010)~\cite{Mainz2010} data from spectrometer B with $Q^2~<~0.02$~GeV$^2$. We used $P_{2,0}$ and created functions with $c_0 = 1$ and 500 evenly spaced constants with $\sqrt{6 c_1} \hbar c~=~R_E$ in [0.750,~0.950]~fm and $c_2$ in [-30.0, 30.0]~GeV$^{-4}$ for a total of $500^2$ parameterizations.

\begin{figure}[h!]
\includegraphics[scale=0.45]{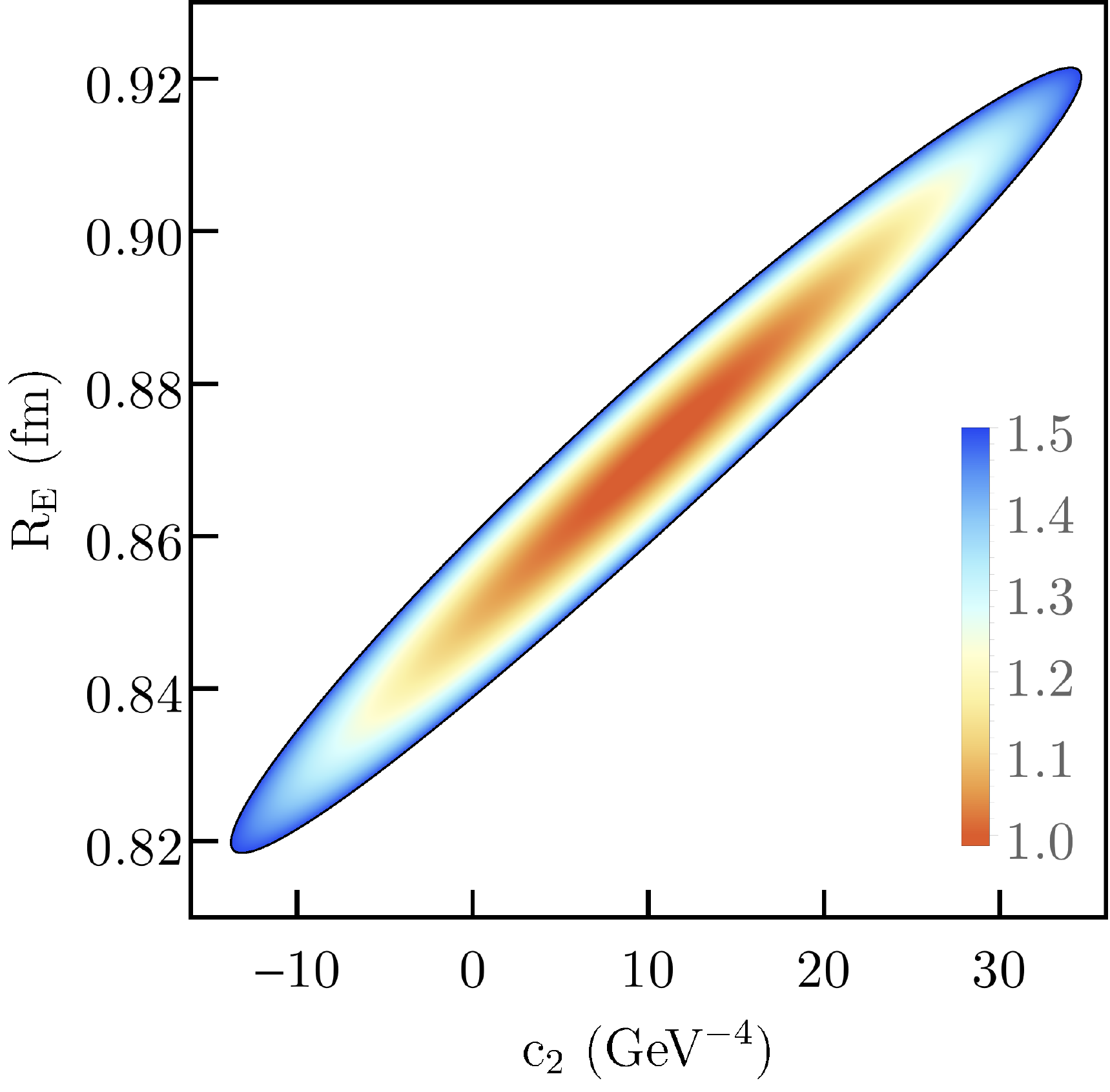}
\caption{(Color online). Two-dimensional $\chi^2/\text{dof}$ surface for the low-$Q^2$ Mainz (2010) data fit to $P_{2,0}$ as a function of the parameters $R_E$ and $c_2$ with the overall normalization $c_0 = 1$. The $\chi^2$/dof surface indicates a wide range of acceptable radii.}
\label{heat_plot}
\end{figure}

The $\chi^2/$dof for each parameterization is shown in FIG.~\ref{heat_plot} as a function of the proton radius and curvature terms. The $\chi^2$/dof alone implies a wide range of acceptable radii from $\approx$ 0.84 to 0.90~fm. This $\chi^2$ trench arises when increased quadratic curvature is offset by a larger slope at the origin. 

\begin{figure}[h!]
\includegraphics[scale=0.41]{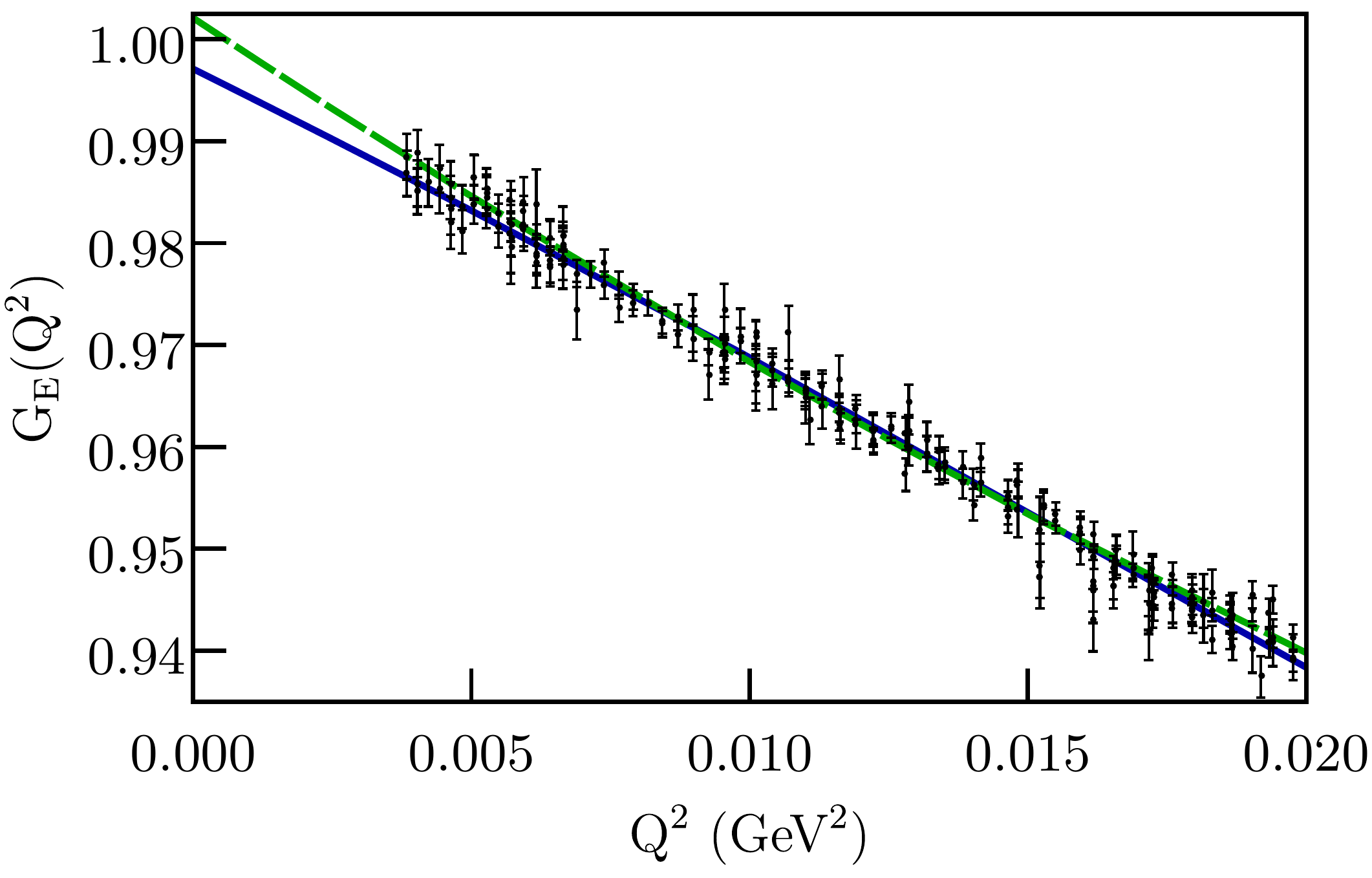}
\caption{(Color online). Three $P_{2,0}$ fits of the Mainz low-$Q^2$ data with low $\chi^2$/dof but with extracted radii differing by more than 0.1~fm. The solid blue line has a radius of 0.800~fm and $\chi^2$/dof~=~1.033, the dotted red line has a radius of 0.865~fm and $\chi^2$/dof = 0.986, and the dashed green line has a radius of 0.920~fm with a $\chi^2$/dof = 1.025. Note that the fits are not constrained to $G_E(0)=1$ because of the normalization constant.} 
\label{varied_good_fits}
\end{figure}

This is demonstrated in another way in FIG. \ref{varied_good_fits}. Three different fits of the $P_{2,0}$ function to the low-$Q^2$ Mainz~(2010) data are pictured. Each fit has a $\chi^2$/dof near unity but their extracted radii vary by more than 0.1~fm. The fits are visually similar above $Q^2~\approx~0.0075$~GeV$^2$ but they differ significantly at lower $Q^2$ where there is no data. These results should be a caution that a good $\chi^2$/dof value alone does not imply an accurate radius extraction.  

Any experimental measurement is only one instance of a distribution of possible values governed by statistical variance. To understand the implications of this variance, we generated a data set with 15 points using the $P_{1,0}$ function with 1\% uncertainty on each point. Each point was then shifted by a random value chosen from a Gaussian distribution with $\sigma$ set by the uncertainty on the point. Then the data set was refit using $P_{1,0}$. This process was repeated 10,000 times. The resulting distribution of $\chi^2$ values, as expected, was a $\chi^2$ distribution with $k$ degrees of freedom, given by
\begin{align}
f(x,k) &= \frac{x^{\frac{k}{2}-1}e^{-\frac{x}{2}}}{2^{\frac{k}{2}}\Gamma\left( \frac{k}{2} \right)},
\end{align}
in which $\Gamma$ denotes the Euler-gamma function and $x > 0$.

\begin{figure}[h!]
	\includegraphics[scale=0.41]{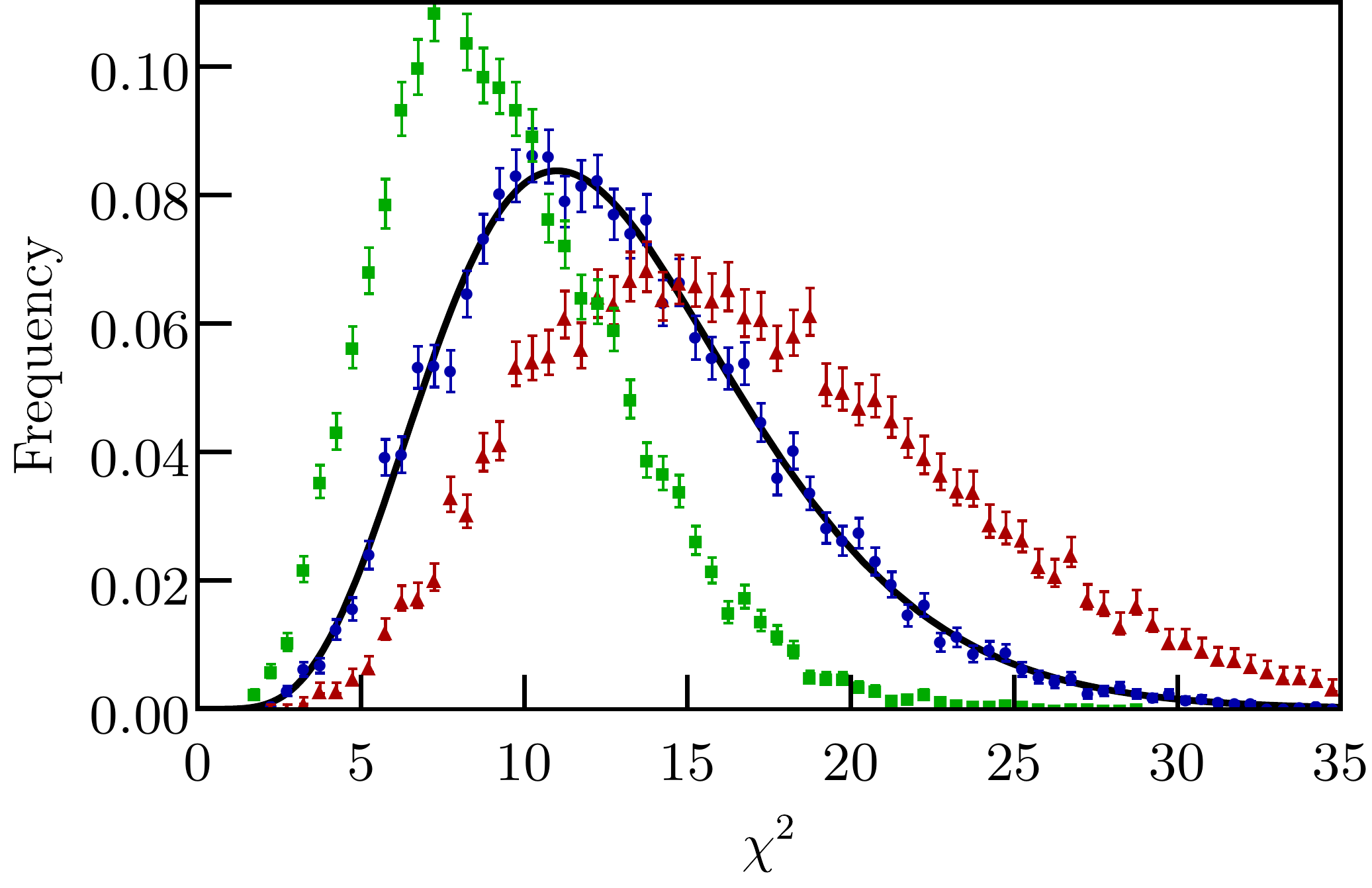}
	\caption{(Color online). Histogram of $\chi^2$ values for data created by allowing the data points to fluctuate within their variance by $\sigma$ (blue circles), $1.15 \sigma$ (red triangles) and $0.85 \sigma$ (green squares). The solid black line is the expected $\chi^2$ distribution.}
	\label{chi2_distros}
\end{figure}

For a data set with 15 points, fit to $P_{1,0}$, we expected $k=13$. FIG.~\ref{chi2_distros} shows the $\chi^2$ distribution for $k=13$ with the pseudodata (blue circles) agreeing well. Despite this agreement for the distribution, it is possible for a single fit to return a $\chi^2$ value much larger or smaller than $\chi^2/\text{dof}~=~1$. A single measured unsatisfactory value for $\chi^2$ from experimental data does not necessarily indicate a poor estimate of a fitted parameter.

Over- or under-estimated point-to-point errors introduce a potential problem. We studied this by repeating the above procedure, but instead shifted the points by a random value chosen from a Gaussian distribution with width equal to $1.15 \sigma$ (red triangles, representing quoted error bars that are too small) and $0.85 \sigma$ (green squares, representing quoted error bars that are too large). The results for these distributions of $\chi^2$ are also pictured in FIG.~\ref{chi2_distros}. If the statistical uncertainties on the data points are not representative of true Gaussian statistics, it may not be accurate to say that a poorly reconstructed $\chi^2$ distribution indicates a bad extraction.

For each fit in this paper we use the quantity $k'/k$ as an indicator of the statistical reliability of the experimental error estimates on the data points. Here $k$ is the number of degrees of freedom of the fit and $k'$ is the number of degrees of freedom obtained by fitting the distribution of $\chi^2$ values from the 10,000 fits. In Monte Carlo tests we found that the ratio $k'/k$ was proportional to $(\sigma'/\sigma)^2$, where $\sigma'$ corresponded to a rescaled value of the experimental uncertainties. Thus, the $\chi^2$/dof of a fit can be made equal to unity by scaling the error bars on the data by $\sqrt{k'/k}$. If the experimental uncertainties are over-estimated, $k'/k<1$ and if the uncertainties are under-estimated, $k'/k>1$. Although this does not change the central values for $R_E$, it does affect the uncertainty quoted for $R_E$. We include this effect as a multiplicative factor on the variance for each extracted radius.

\subsection{Fitting Algorithm}
We embarked on a analysis of archival form factor data to better understand the current limitations on extracting nuclear radii. Our goal was to create a consistent procedure that worked to simultaneously minimize both the statistical uncertainty and possible bias stemming from the chosen fit model. There is potential ambiguity in the choice of what $Q^2$ range to use from each published data set and so we evaluated the statistical uncertainty and estimated the bias for each possible contiguous subset of the overall data. For each data set, we took the published data at face value and accepted reported uncertainties as statistically distributed. Parameters for each fit were chosen by minimizing Eq.~\ref{chi2function} via the MIGRAD algorithm implementation in MINUIT of ROOT~\cite{Minuit}. We required the fit to converge for all iterations of the radius extraction procedure described below. 

\begin{figure}[h!]
	\includegraphics[scale=0.38]{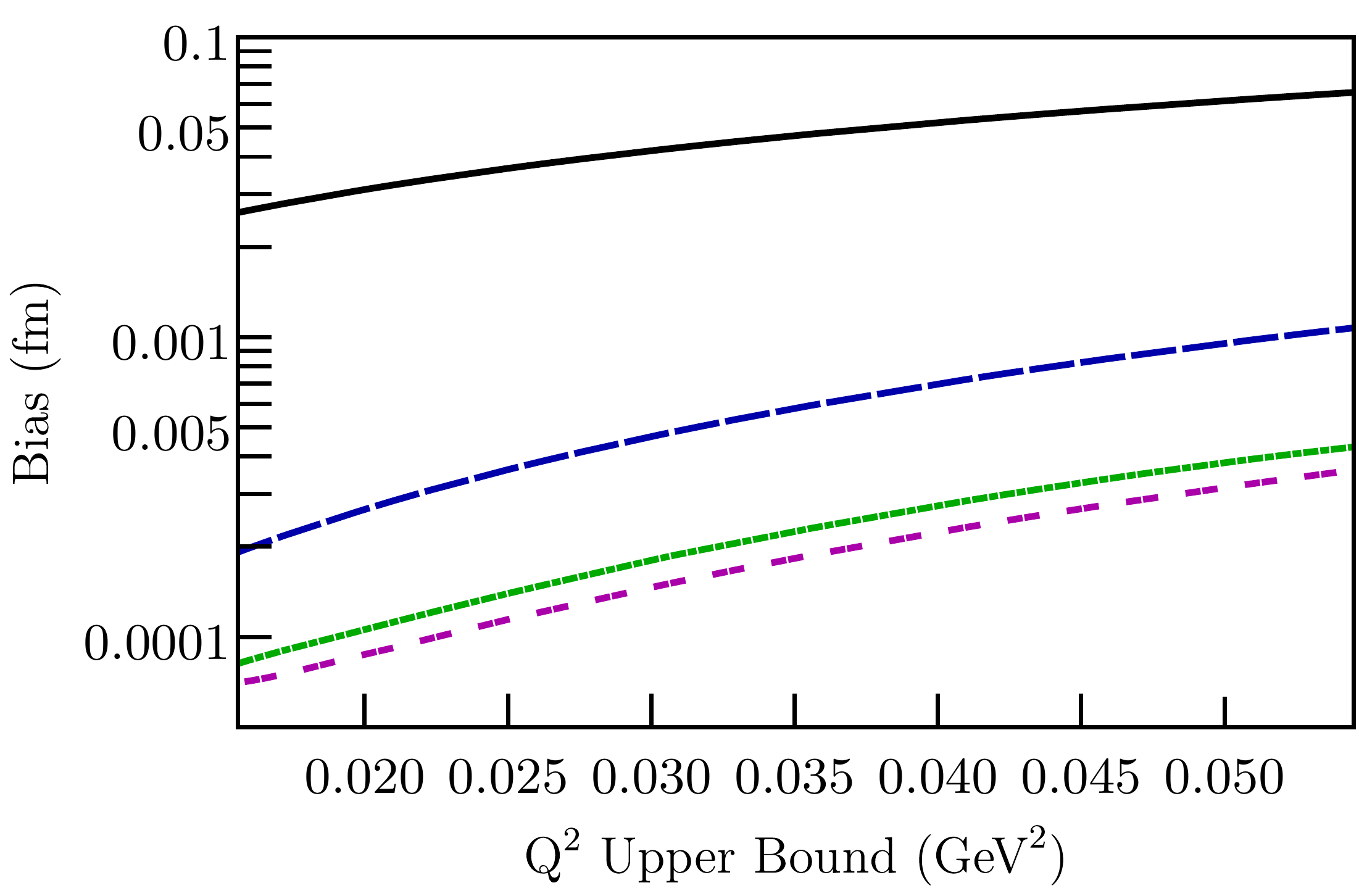}
	\caption{(Color online). The systematic bias of several fit functions over the range in possible $Q^2$ upper bounds for the Mainz (1980) data set. The functions with one parameter, $P_{1,0}$ (solid black line) and $P_{0,1}$ (dotted red line), show less flexibility and a corresponding increase in bias than do the functions with two parameters, $P_{2,0}$ (dashed blue line), $P_{0,2}$ (dash-dotted green line) and ${CF}_2$ (double dashed magenta line).}
	\label{systematic_bias}
\end{figure}

The bias of each fit form at each possible $Q^2$ upper bound of the examined data set was estimated. This was done by generating 10,000 sets of pseudodata with the generating functions discussed in Section \ref{models}, over the range of $Q^2$ values present in the examined data set. These pseudodata were then fit with the proposed fit forms from their first point to the $i$th point, and then the $i$th$+1$ point, and so on, until the entire data set had been fit. The value $i$ was the minimum number of data points necessary to fit each function (i.e., greater than the number of free parameters). The bias was determined to be the rms value of the difference between the input radii in the generating functions and the extracted radii from the fit functions. The results of this procedure for some of the fit functions applied to the Mainz (1980)~\cite{Mainz1980} data set are pictured in FIG.~\ref{systematic_bias}. This process allowed us to estimate the bias of each potential fit as a function of the $Q^2$ upper bound. The functions with fewer parameters produce systematically higher biases than the functions with more parameters. This procedure for estimating the bias is similar to that of Ref. \cite{Yan:2018bez}.

\begin{figure}[h]
	\includegraphics[scale=0.38]{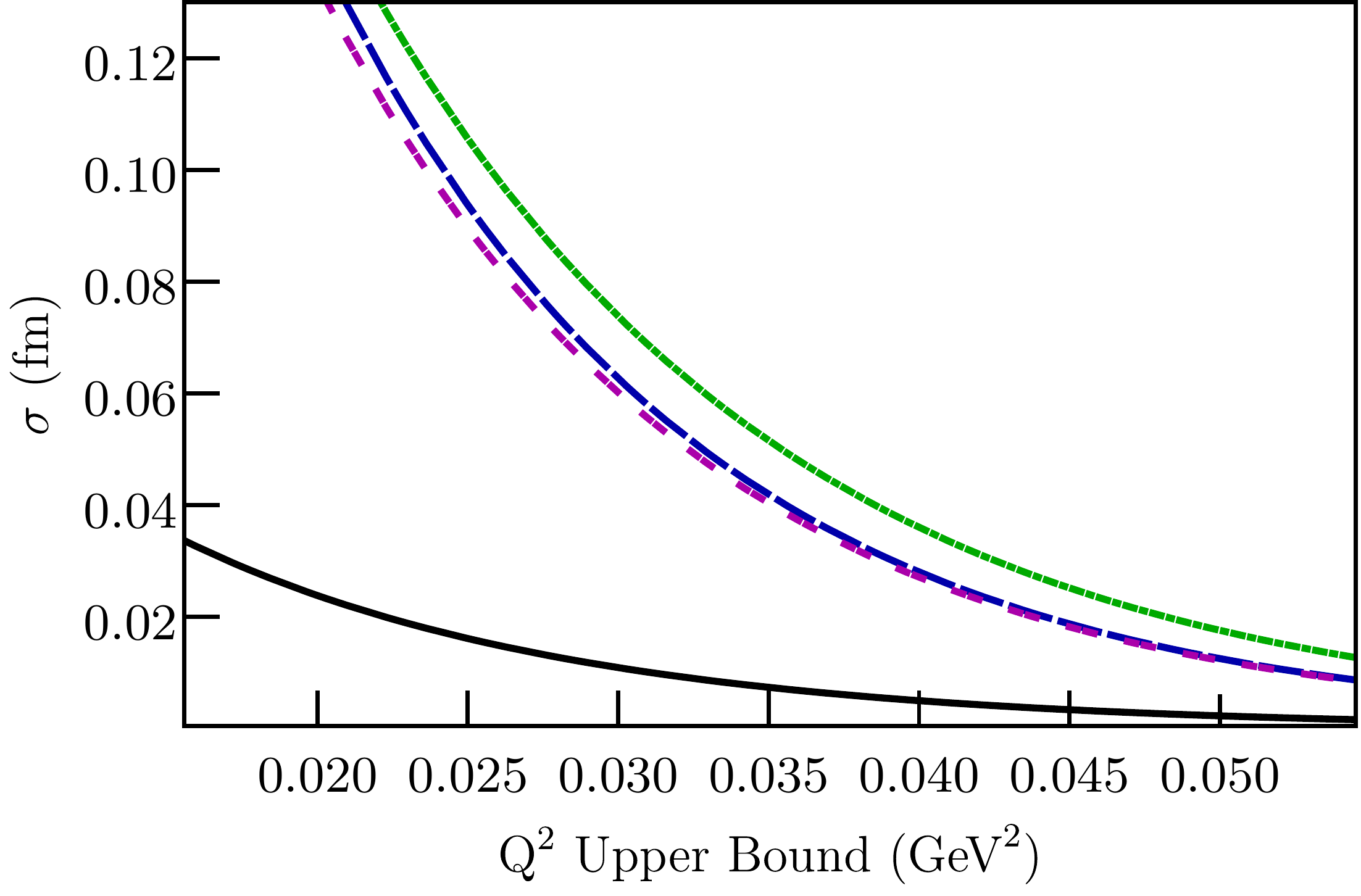}
	\caption{(Color online). The statistical uncertainty of each fit function over the range in possible $Q^2$ upper bounds for the Mainz (1980) data set. The functions with one parameter, $P_{1,0}$ (solid black line) and $P_{0,1}$ (dotted red line), produce smaller uncertainties than the functions with two parameters, $P_{2,0}$ (dashed blue line), $P_{0,2}$ (dash-dotted green line) and ${CF}_2$ (double dashed magenta line).}
	\label{uncertainty}
\end{figure}

To determine the statistical uncertainty on the extracted radius, the individual points in the data set were randomly shifted based on their uncertainty and the data were refit. This process was repeated for each fit function and $Q^2$ upper bound and the statistical uncertainty in each case was taken to be the rms variation of the 10,000 extracted radii. The statistical uncertainty as a function of $Q^2$ upper bound for a number of functions fit to the Mainz (1980) data set is pictured in FIG. \ref{uncertainty}. 

More complex functions generally produce larger variances but smaller biases and vice versa. The best fit function and $Q^2$ upper bound will minimize the combination of variance and bias. We combined the statistical uncertainty and bias in quadrature in order to estimate the total uncertainty on the extracted radius. The radius chosen for each data set was extracted using the particular function and $Q^2$ upper bound (not necessarily the entire range of published data) that had the smallest combination of bias and variance of the candidates whose functional forms converged for every iteration.

\section{Historical Data}
\subsection{Proton}

We examined 5 early $ep$ scattering data sets. The first, Hand \emph{et al}. (1963)~\cite{Hand1963} gave a review of scattering data before 1963. We use the published values of $G_E$ below 0.116~GeV$^2$~(3~fm$^{-2}$) for a loose definition of ``low-Q$^2$." These data were comprised of points from 5 different experiments and the $G_E$ values were derived from a Rosenbluth separation of the cross section data for $Q^2 \geq 0.078$~GeV$^2$ and by implicit assumption that $G_M = \mu G_E$ for values with $Q^2 < 0.078$~GeV$^2$. The next two data sets were Yerevan (1972)~\cite{Yerevan1972} and Saskatoon (1974)~\cite{Saskatoon1974} whose published $G_E$ values were calculated from a Rosenbluth separation. The fourth data set was Mainz (1975)~\cite{Mainz1975} which included published $G_E$ values derived via Rosenbluth separation from experimental cross sections listed in that work and in Ref. \cite{Mainz1974}. The final 20th century $ep$ data set was Mainz (1980)~\cite{Mainz1980} which contains values for $G_E$ calculated assuming $G_M = \mu G_E$. 

The Mainz (2010) cross section data is given as a ratio of the measured cross section, $\sigma$, to the dipole cross section, $\sigma_D$,
\begin{equation}
\frac{\sigma}{\sigma_D} = \frac{\epsilon G_E^2 + \tau G_M^2}{\epsilon G_D^2 + \tau \mu_p^2 G_D^2}.
\end{equation}
From this,
\begin{equation}
G_E = G_D \left( \frac{\sigma}{\sigma_D} \right)^{1/2} \left[ 1 + \tau \mu_p^2 \frac{G_M^2 / (\mu_p G_E)^2 - 1}{\epsilon + \tau \mu_p^2} \right]^{-1/2}.
\end{equation}
We used the $G_E/G_M$ ratio obtained from recoil polarization experiments in order to extract $G_E$ for the Mainz (2010) data set. JLab data \cite{Jones:1999rz,Gayou:2001qd,Puckett:2010ac,Punjabi:2005wq,Puckett:2011xg,Ron:2011rd} indicate that for low-$Q^2$ data, 
\begin{equation}
\mu_p \frac{G_E}{G_M} \approx 1 - \frac{Q^2}{8~\text{GeV}^2}.
\end{equation}
The effect of slightly altering the value of 8~GeV$^2$ over the range 5 to 15 GeV$^2$ was studied in Ref \cite{smallRadii2} and was found to change the extracted radius by a negligible amount when compared to other uncertainties. 

Table \ref{proton_table1} lists the results for $R_E$ from our fits to the historical data sets. The 20th century data sets span the range of $R_E$ = 0.80 to 0.90 fm. We attribute this effect to the long and deep $\chi^2$ trench (see FIG. \ref{heat_plot}) which indicates that the linear and quadratic terms at low-$Q^2$ are not independently constrained. The Mainz (2010) data is much more precise than the other sets and our extracted radius, 0.841(1)~fm, is in precise agreement with the muonic Lamb shift value.

\begin{center}
	\begin{table}[h!]
		\begin{tabular}{|c |c| c| c| c|} 
			\hline
			Data Set & $R_E$ (fm) & $F(Q^2)$ & $Q^2$ Range (GeV)$^2$ & $k'/k$  \\ 
			\hline
			Hand \emph{et al}.  (1963) & 0.809(30) & $P_{0,1}$ &(0.011, 0.062)& 1.75 \\
			\hline
			Yerevan (1972) & 0.820(19) & $P_{0,1}$ & (0.008, 0.020) &0.71 \\
			\hline
			Sasktoon (1974) & 0.864(24) & $P_{0,1}$ & (0.005, 0.031)& 1.75\\ 
			\hline
			Mainz (1975) & 0.849(36) & $P_{2,0}$&(0.014, 0.123) &0.77 \\
			\hline
			Mainz (1980) &0.884(19) &  $P_{0,1}$ & (0.005, 0.025) & 1.50\\
			\hline
			Mainz (2010) &0.841(1) & ${CF}_{4}$ & (0.004, 0.326) &3.31\\			
			\hline
		\end{tabular}
		\caption{The proton radius values determined by minimizing the combination of statistical uncertainty and bias for each data set.}
		\label{proton_table1}
	\end{table}
\end{center}

\begin{figure}[h!]
\includegraphics[scale=0.41]{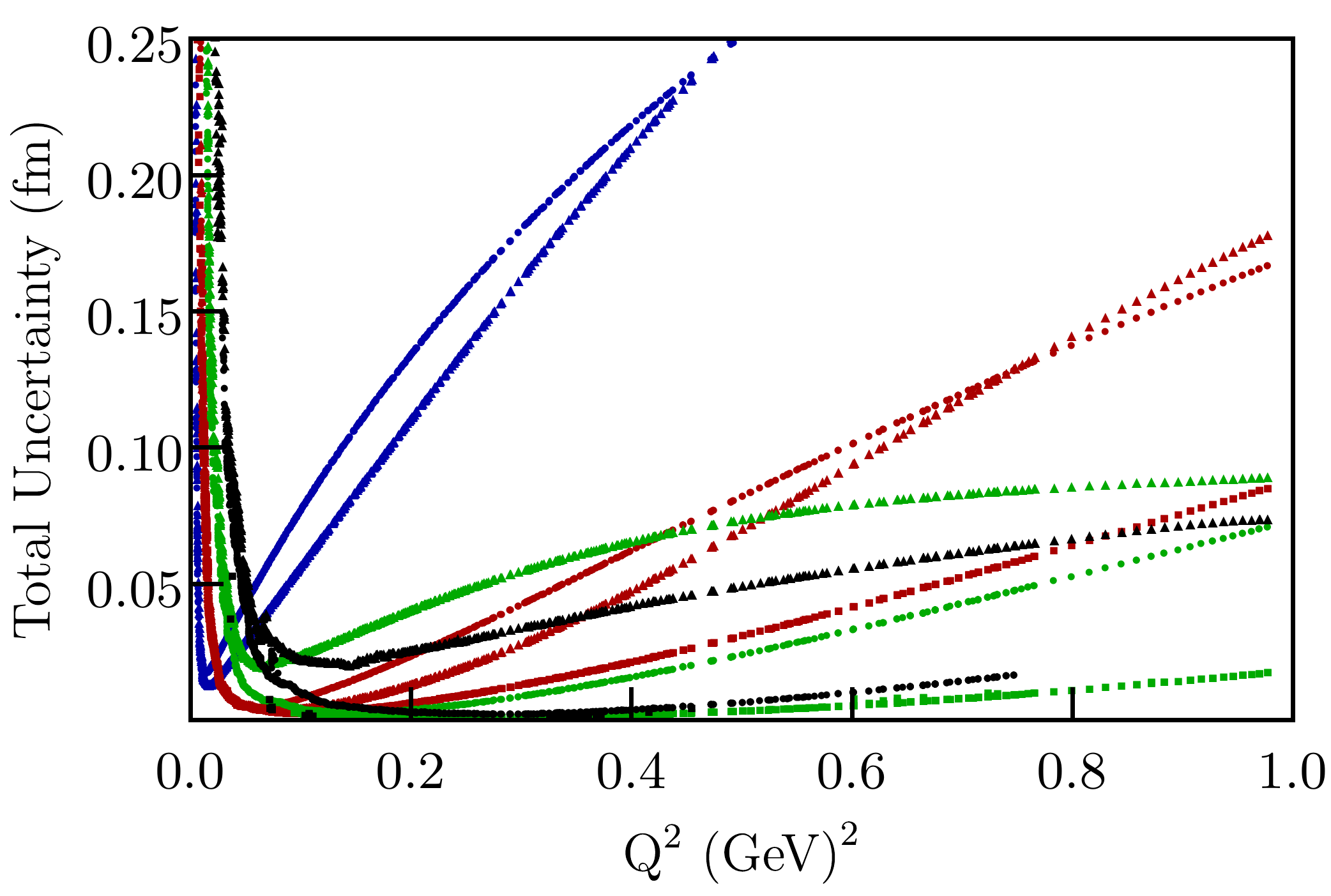}
\caption{(Color online). The total uncertainty (combined statistical and bias) in the Mainz (2010) fits as a function of maximum $Q^2$ for the one, two, three and four parameter models in blue, red, green and black respectively and functions of the form $P_{i,0}$, $P_{0,i}$ and ${CF}_i$ represented by circles, triangles and squares, respectively. }
\label{total_uncertainties}
\end{figure}

\subsection{Model Dependence} 
To further investigate possible uncertainties coming from the fitting process, we examined the total uncertainty (combined statistical and bias) as a function of maximum $Q^2$ for each fit function applied to the Mainz (2010) data set. FIG. \ref{total_uncertainties} shows that there are many fit options with similar uncertainties below, for example, a total uncertainty of 0.05~fm. Consequently, it is necessary to understand the model dependence of these fits.

In order to resolve fit ambiguities, we attempted to include the contributions from every possible fit that converged. This was done by evaluating the quantity
\begin{align}
P(R_E) &= \sum\frac{1}{{\sigma'}^2} \mathcal{N}  \left( \mu, \sigma' \right),
\end{align}
in which the summation runs over all eleven fit functions and maximum $Q^2$ combinations and $\mathcal{N}$ is the Gaussian distribution, $\mu$ is the mean of the 10,000 fits and $\sigma'$ is the statistical uncertainty of those fits scaled by the $\sqrt{k'/k}$ ratio as determined from the distribution of $\chi^2$ values. In order to avoid issues coming from the build-up of common uncertainties, we limited the summation to fits whose estimated bias values were less than 0.008~fm. This number was chosen because it corresponds to a 5$\sigma$ difference between the 0.84 and 0.88~fm radius values. We studied the effect of varying this limit (discussed below) and found little effect on the results. In $P(R_E)$ the precise, unbiased fits should cluster around the most likely radius and subtly biased fits (within the upper bound of acceptable bias) should fill out the ends of the distribution.

Model error was calculated for every radius value in the previous section by determining the central range required to cover 0.683\% (1$\sigma$) of the distribution, $P(R_E)$. This is generally not explicitly Gaussian, but serves as a rough estimate of the model uncertainty due to functional form and $Q^2$ range. Each $ep$ data set was reevaluated and the results are given in TABLE \ref{proton_table}. 

\begin{figure}[h!]
\includegraphics[scale=0.41]{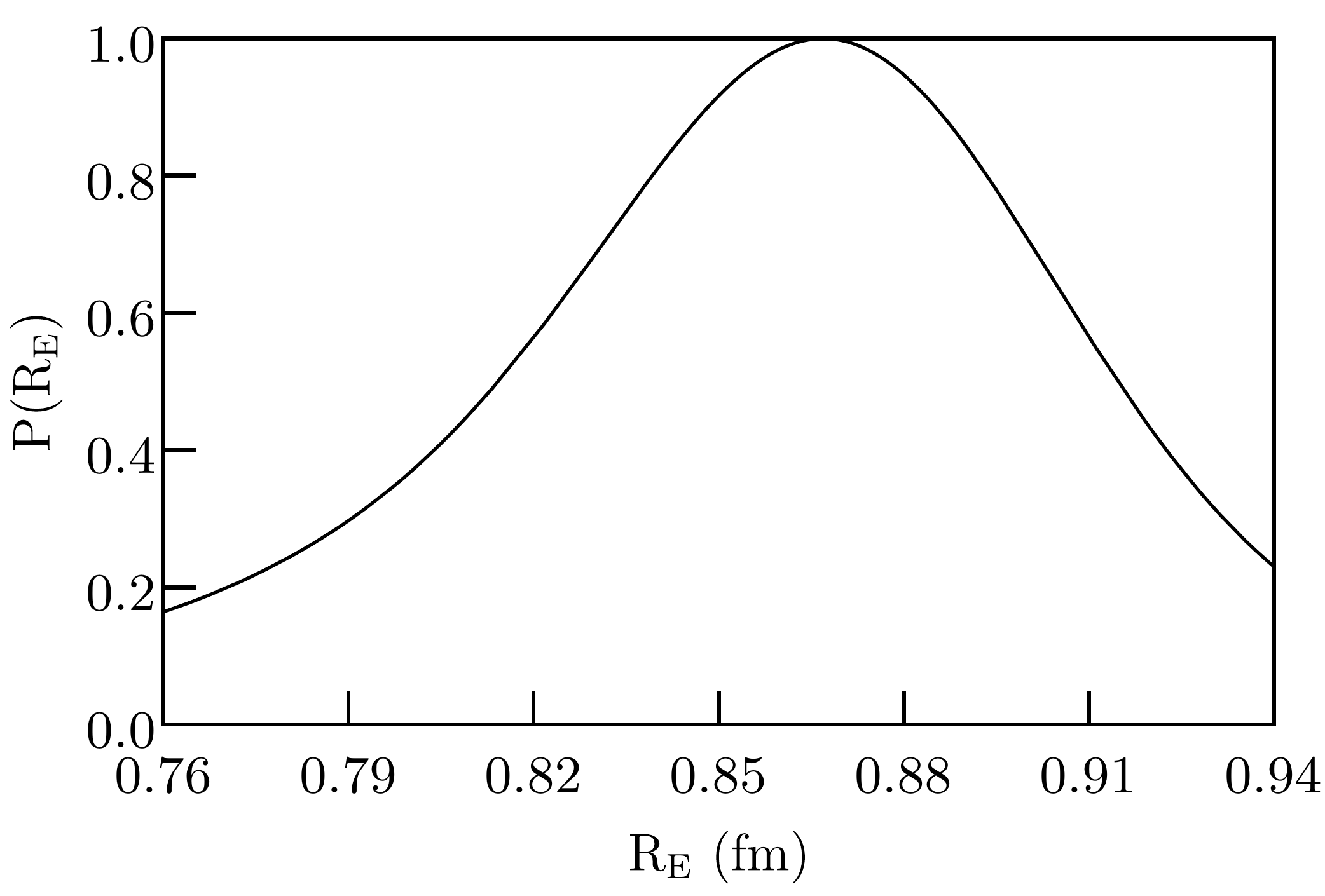}
\caption{$P(R_E)$ for the Mainz (1980) data set. The peak of the distribution is 0.867~fm and the model error is 0.024~fm.}
\label{mainz80_gaussians}
\end{figure}

$P(R_E)$ for the Mainz (1980) data set is pictured in FIG. \ref{mainz80_gaussians} as an example. Based on the distribution, we assign the Mainz (1980) data set a model uncertainty of 0.024~fm and combine that in quadrature with the previously quoted uncertainty to arrive at a radius value of 0.884(31)~fm. This uncertainty includes variance, estimated systematics via study of the $\chi^2$ distribution and model uncertainty. This result has larger uncertainties than those traditionally reported because of this ambiguity in the appropriate way to extrapolate to $Q^2=0$. As discussed previously, smaller or larger values of proton radius extractions can be attributed to fits resolving the linear/quadratic ambiguity in separate ways as shown in FIG.~\ref{heat_plot}.

\begin{figure}[h!]
\includegraphics[scale=0.41]{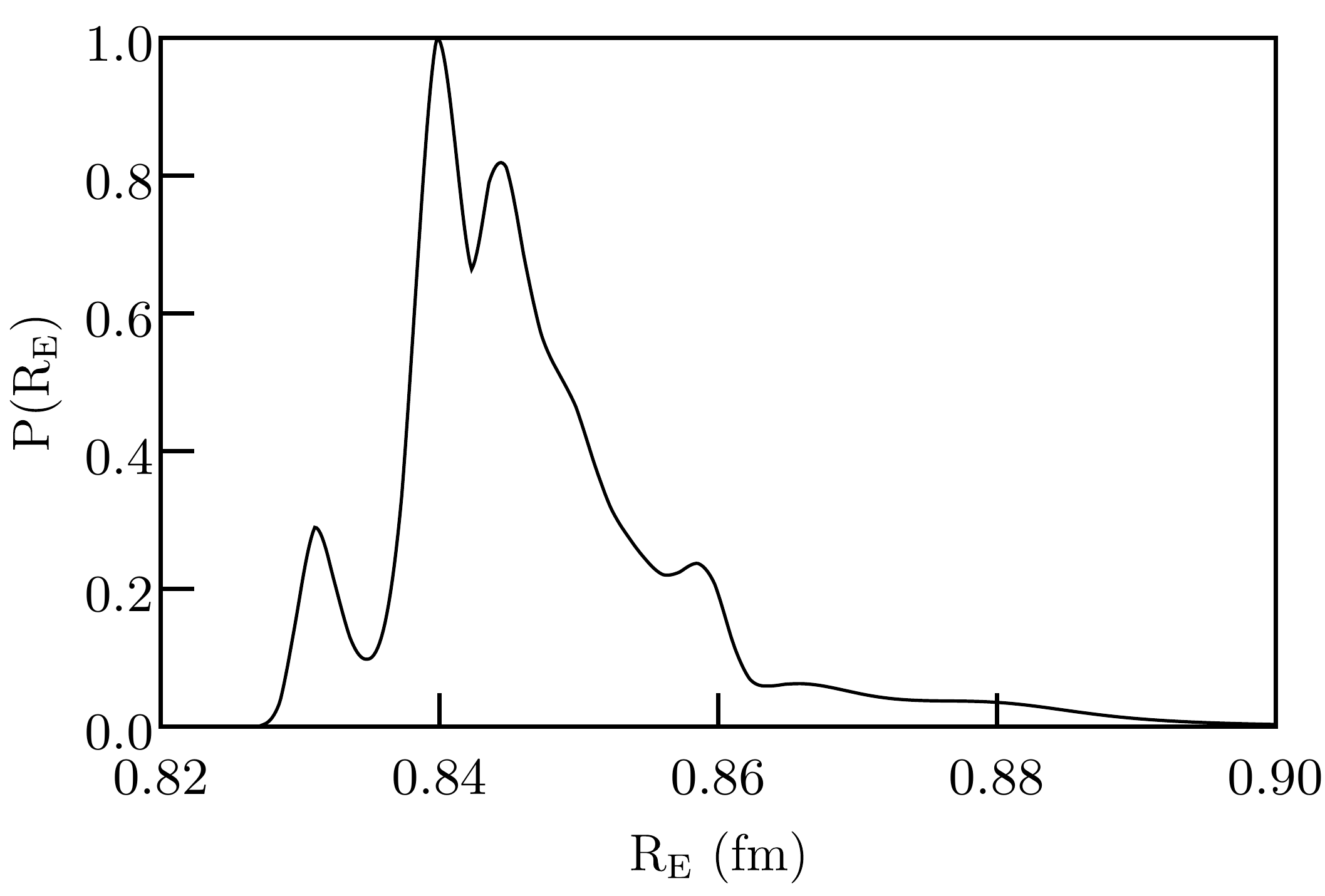}
\caption{$P(R_E)$ for the Mainz (2010) data set. The most likely value of the distribution is 0.840~fm and the model error is 0.004~fm. The smallest peak at around 0.831~fm is comprised of polynomial fits, mostly $P_{4,0}$, which slightly underestimate the radius. The two larger peaks, at around 0.840 and 0.844~fm are primarily comprised of $CF_3$ and $CF_2$ fits respectively. The asymmetric distribution to the right comes from various models that slightly overestimate the radius, such as inverse polynomials.}
\label{mainz10_gaussians}
\end{figure}

The historical 20th century data does not have the precision to convincingly differentiate between the large and small radii. The Mainz (2010) data set, whose $P(R_E)$ is pictured in FIG. \ref{mainz10_gaussians}, is much more precise. The distribution shown in FIG. \ref{mainz10_gaussians} is non-Gaussian and the 1-$\sigma$ (68.3\%), 2-$\sigma$ (95.4\%) and 3-$\sigma$ (99.7\%) ranges are 0.841-0.849~fm, 0.832-0.871~fm and 0.829-0.892~fm respectively. 

We determined a proton radius from the Mainz (2010) data set of 0.841(4)~fm, where the central value comes from the fitting algorithm described above and the majority of the error comes from the estimate of the model dependence. This value is in good agreement with the muonic-hydrogen value, 0.84087(39)~fm. We studied the effect of altering the maximum allowed bias in $P(R_E)$ from 0.002 to 0.012~fm and the found the mean value and widths remained stable within 0.002~fm. There is little to no effect on the estimate of the model error coming from $P(R_E)$ when the maximum bias value is altered in these ranges.

\begin{center}
		\begin{table}[h!]
			\begin{tabular}{|c |c |c| } 
				\hline
				Data Set & Modeling Error (fm) & $R_E$ (fm)  \\ 
				\hline
				Hand \emph{et al}.  (1963) & 0.074 & 0.809(80)  \\
				\hline
				Yerevan (1972) & 0.067 &0.820(70)  \\
				\hline
				Sasktoon (1974) & 0.057 &0.863(62)  \\ 
				\hline
				Mainz (1975) & 0.032 & 0.849(48)  \\
				\hline
				Mainz (1980) & 0.024 &0.884(31)  \\
				\hline
				Mainz (2010) &0.004 &0.841(4) \\			
				\hline
			\end{tabular}
			\caption{The final results for $R_E$ after studying the model dependence of the radius extracted from each $ep$ scattering data set.}
			\label{proton_table}
		\end{table}
\end{center}

\begin{figure}[h!]
	\includegraphics[scale=0.37]{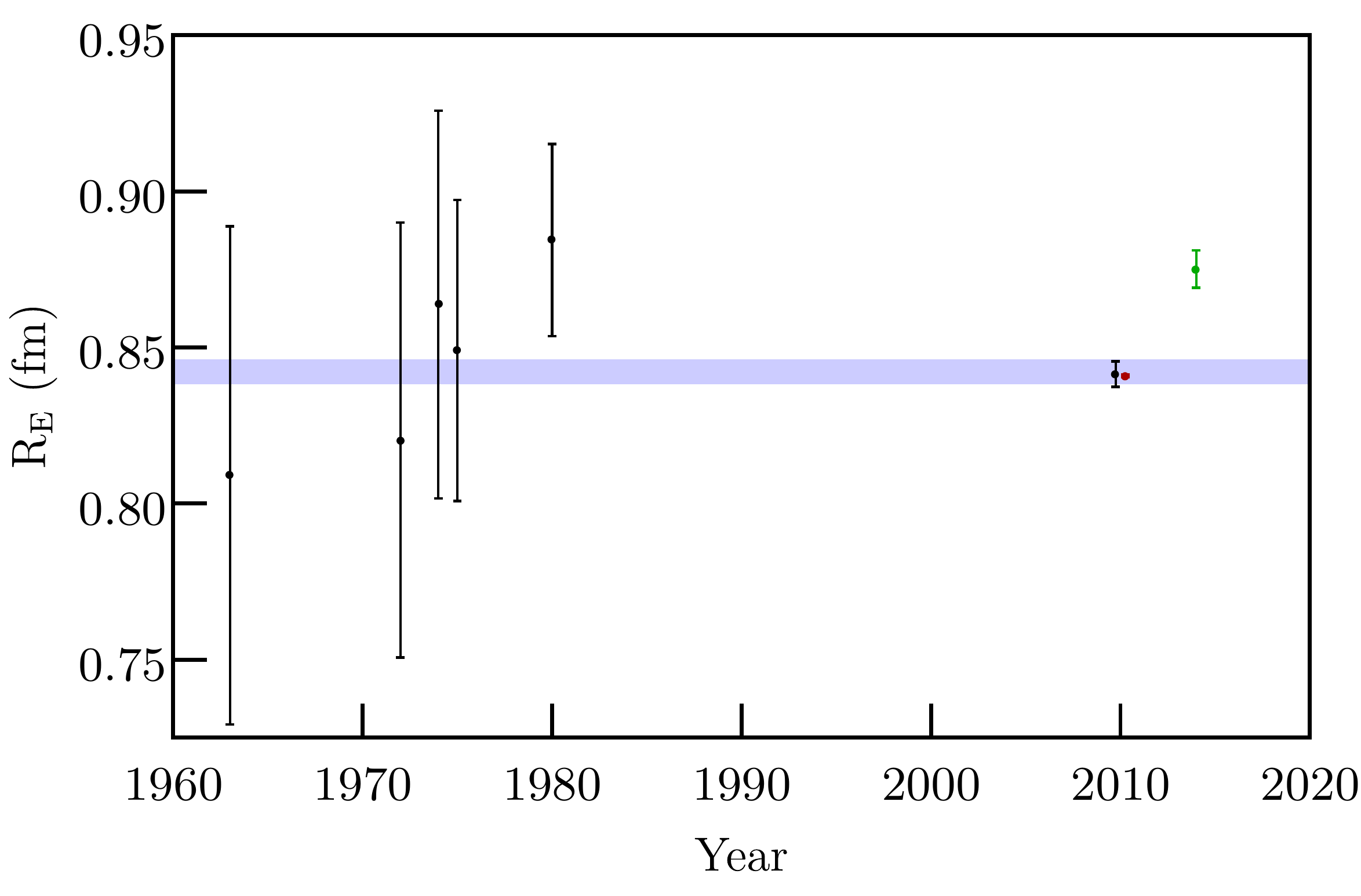}
	\caption{(Color online). The proton radii extracted from each $ep$ scattering data set (black), compared to the 2010 muon spectroscopy result, 0.84087(39)~fm (red) and the 2014 CODATA result, 0.875(6)~fm (green). Our weighted average, 0.842(4)~fm, is shown as a blue band.}
	\label{progression_of_radii}
\end{figure}

In FIG. \ref{progression_of_radii} we show proton radius results from each of the examined $ep$ scattering data sets. Our weighted average of 0.842(4)~fm is dominated by the Mainz (2010) result.

\subsection{Deuteron Scattering}
For the deuteron we focused on fitting $A(Q^2)$ from two low-$Q^2$ data sets: Simon \emph{et al}.~(1981)~\cite{Simon1980}, which published $A(Q^2)$ values calculated directly from the cross section for $Q^2 \leq 0.04$~GeV$^2$ and via Rosenbluth separation for $Q^2 \geq 0.04$~GeV$^2$ and Platchkov \emph{et al}.~(1990)~\cite{Platchkov1990} where we used published $A(Q^2)$ values below $Q^2~=~0.116$~GeV$^2$ that were calculated by subtracting $B(Q^2)$ contributions from the cross section using a fit of existing $B(Q^2)$ measurements.

For the Simon (1981) data set we find the minimum combination of statistical uncertainty and bias to come from the ${CF}_3$ fit out to $Q^2$ = 0.082~GeV$^2$ which corresponds to a radius of 2.092(19)~fm, in which the value corresponding to model uncertainty is included. It was difficult to find an appropriate model for the Platchkov (1990) data set. The curvature in the deteuron form factor is much larger than the proton and there is significant freedom in the unmeasured low-Q$^2$ region. The best fit for the Platchkov (1990) data was the $P_{3,0}$ function with a maximum $Q^2$ of 0.071~GeV$^2$ and a radius of 1.796(194)~fm. Both of these values indicate a small radius, although the uncertainty on the Platchkov (1990) result makes extracting anything specific impossible. Our final result for the deuteron radius, 2.092(19)~fm, is then completely dominated by the Simon (1981) data.

\section{Conclusions}
Using a consistent and comprehensive fitting approach we have arrived at a proton radius measured from $ep$ scattering of $0.842(4)$~fm which is consistent with the muonic hydrogen value of 0.84087(39)~fm. We found a deuteron radius from $ed$ scattering of 2.092(19)~fm which indicates a smaller radius, similar to the value from muonic deuterium, 2.1256(8)~fm, but with large errors. We stress that systematic uncertainties have not been estimated in this analysis beyond a scaling of the uncertainties based on studies of the $\chi^2$ distribution for a particular fit.

Furthermore, our studies demonstrate how the discrepancy between small and large radii arises in fits to elastic scattering data. Depending on how the linear/quadratic ambiguity is resolved, reasonable fits can yield radii from 0.84 to 0.88 fm. Our analysis indicates that a smaller result is more likely, but precise new experiments that extend to lower~$Q^2$ are needed. Several upcoming elastic-scattering efforts should make valuable contributions toward solving the proton radius problem by helping to resolve the linear/quadratic ambiguity. No $ed$ scattering data set exists with comparable statistics to the Mainz (2010) $ep$ experiment and so the uncertainty on the deuteron radius from $ed$ scattering remains large. Upcoming results from a Mainz A1 experiment should eclipse previous $ed$ studies and help shed light on the deuteron radius problem.

\begin{acknowledgments}
	We thank Carl Carlson and Douglas Higinbotham for useful discussions and the Mainz Institute for Theoretical Physics for its hospitality and support during a portion of these studies. This work was also supported by the Department of Energy under grant DE-FG02-96ER41003.
\end{acknowledgments}

\leavevmode


\begin{thebibliography}{99}

%%\cite{Carlson:2015jba}
%\bibitem{Carlson:2015jba} 
%  C.~E.~Carlson,
%  %``The Proton Radius Puzzle,''
%  Prog.\ Part.\ Nucl.\ Phys.\  {\bf 82}, 59 (2015)
%  doi:10.1016/j.ppnp.2015.01.002
%  [arXiv:1502.05314 [hep-ph]].
%  %%CITATION = doi:10.1016/j.ppnp.2015.01.002;%%
%  %77 citations counted in INSPIRE as of 17 Aug 2017
  
    \bibitem{Hand1963}
  L. N. Hand, D. G. Miller and Richard Wilson.
  % ``Electric and Magnetic Form Factors of the Nucleon.''
  Review of Modern Physics, {\bf 35}, no 2, pp. 335-349 (1963).
  %%CITATION = 10.1103/RevModPhys.35.335 %%
  
  \bibitem{Yerevan1972}
  Yu. K. Akimov, \emph{et al}.
  %``Electron Scattering by Protons at Small Angles''
  Soviet Physics JEPT, {\bf 35}, no 4. (1972).
  
  \bibitem{Saskatoon1974}
  J. J. Murphy, II, Y. M. Shin and D. M. Skopik.
  % ``Proton form factors from 0.15 to 0.79 fm$^{-2}$.''
  Phys.\ Rev.\ C {\bf 9}, no 6, (1974).
  
  \bibitem{Mainz1975}
  F Borkowski, G. G. Simon, V. H. Walhter and R. D. Wendling.
  %``Electromagnetic form factors of the proton at low four-momentum transfer.''
  Nuc.\ Phys.\ {\bf B93}, Issue 3, pp 461-478 (1975).
  
    \bibitem{Mainz1974}
    F Borkowski, P. Peuser, G. G. Simon, V. H. Walhter and R. D. Wendling.
   %``Electromagnetic form factors of the proton at low four-momentum transfer.''
   Nuc.\ Phys.\ {\bf A222}, pp 269 - 275 (1974)
  
  \bibitem{Mainz1980}
  G. G. Simon, Ch. Schmitt, F. Borkowski and V. H. Walhter.
  %``Absolute electron-proton cross sections at low momentum transfer measured with a high pressure gas target system.''
  Nuc.\ Phys.\ {\bf A333}, pp. 381-391, (1980).
  
  \bibitem{Mainz2010}
  J. C. Bernauer \emph{et al}. [A1 Collaboration], Phys. Rev. Lett. 105
  (2010) 242001. [arXiv:1007.5076 [nucl-ex]].
 
  \bibitem{hyd_lamb1}
  M. G. Boshier \emph{et al}. Phys. Rev. A {\bf40}, 6169 (1989).
  
  \bibitem{hyd_lamb2}
  M. Weitz \emph{et al}. Phys. Rev. Lett, {\bf72}, 328 (1994).
  
  \bibitem{hyd_lamb3}
  D. J. Berkeland, E. A. Hinds and M. G. Boshier. Phys. Rev. Lett, {\bf75}, 2470 (1995).
  
  \bibitem{hyd_lamb4}
  S. Bourzeix, B. de Beauvoir, F. Nez, M. D. Plimmer, F. de Tomasi, L. Julien, F. Biraben, and D. N. Stacey.
  %``High Resolution Spectroscopy of the Hydrogen Atom: Determination of the 1S Lamb Shift''
  Phys. Rev. Lett. {\bf76}, Number 3, (1996)
  
    \bibitem{hyd_lamb5}
  Th. Udem, A Huber, B. Gross, J. Reichert, M. Prevedelli, M. Weitz and T. W. H\"{a}nsch.
  %``Phase-Coherent Measurement of the Hydrogen 1S-2S Transition Frequency with an Optical Frequency Interval Divider Chain''
  Phys. Rev. Lett. {\bf79}, Number 14, (1997).
  
 
  
  \bibitem{Pohl2010}
   R. Pohl \emph{et al.}, Nature {\bf 466} (2010) 213.
   
   \bibitem{Antognini2013}
   A. Antognini, F. Nez, K. Schuhmann, F. D. Amaro, F. Biraben, J. M. R. Cardoso,
	D. S. Covita and A. Dax \emph{et al}., Science {\bf 339}, 417 (2013).
	
	  \bibitem{CODATA}
  P. J. Mohr, B. N. Taylor, and D. B. Newell, Rev. Mod. Phys. {\bf 84}, 1527 (2012), 1203.5425.
  
%\cite{Lorenz:2012tm}
\bibitem{smallRadii00} 
  I.~T.~Lorenz, H.-W.~Hammer and U.~G.~Meissner,
  %``The size of the proton - closing in on the radius puzzle,''
  Eur.\ Phys.\ J.\ A {\bf 48}, 151 (2012)
  doi:10.1140/epja/i2012-12151-1
  [arXiv:1205.6628 [hep-ph]].
  %%CITATION = doi:10.1140/epja/i2012-12151-1;%%
  %80 citations counted in INSPIRE as of 25 Apr 2018
		
	\bibitem{smallRadii1}
	 I. T. Lorenz and U.-G. Meiner, Phys. Lett. B737, 57 (2014), arXiv:1406.2962 [hep-ph].
	 
	 \bibitem{smallRadii2}
	 K. Griffioen, C. Carlson, and S. Maddox, Phys. Rev. C93, 065207 (2016), arXiv:1509.06676 [nucl-ex].
	 
	 \bibitem{smallRadii3}
	 D. W. Higinbotham, A. A. Kabir, V. Lin, D. Meekins, B. Norum, and B. Sawatzky, Phys. Rev. C93, 055207 (2016),
arXiv:1510.01293 [nucl-ex].

	\bibitem{smallRadii4}
	 M. Horbatsch and E. A. Hessels, Phys. Rev. C93, 015204 (2016), arXiv:1509.05644 [nucl-ex].
	 
	 \bibitem{smallRadii5}
	 M. Horbatsch, E. A. Hessels, and A. Pineda, Phys. Rev. C95, 035203 (2017), arXiv:1610.09760 [nucl-th].
	 
	 \bibitem{largeRadii00}
	 Richard J. Hill and Gil Paz, Phys. Rev. D82, 113005 (2010), arXiv:1008.4619 [hep-ph].
	 
	 \bibitem{largeRadii1}
	  J. Arrington and I. Sick, J. Phys. Chem. Ref. Data 44, 031204 (2015), arXiv:1505.02680 [nucl-ex].
	  
	  \bibitem{largeRadii2}
	  G. Lee, J. R. Arrington, and R. J. Hill, Phys. Rev. D92, 013013 (2015), arXiv:1505.01489 [hep-ph].
	  
	  \bibitem{largeRadii3}
	  D. Borisyuk, Nucl. Phys. A843, 59 (2010), arXiv:0911.4091 [hep-ph].
	  
	  \bibitem{largeRadii4}
	  K. M. Graczyk and C. Juszczak, Phys. Rev. C90, 054334 (2014), arXiv:1408.0150 [hep-ph].
	  
	  \bibitem{largeRadii5}
	  J. C. Bernauer \emph{et al}. (A1), Phys. Rev. C90, 015206 (2014), arXiv:1307.6227 [nucl-ex].
	 
	 	  \bibitem{Simon1980}
  G. G. Simon, Ch. Schmitt and V. H. Walhter. Nuc.\ Phys.\ {\bf A364}, pp. 285-296, (1981).
  
  \bibitem{Platchkov1990}
  S. Platchkov, A Amroun, S. Anuffret, \emph{et al}. Nuc.\ Phys.\ {\bf A510}, pp. 740-758, (1990).


	
	%\cite{Pohl:2016glp}
	\bibitem{Pohl:2016glp} 
	R.~Pohl {\it et al.},
	%``Deuteron charge radius and Rydberg constant from spectroscopy data in atomic deuterium,''
	Metrologia {\bf 54}, L1 (2017)
	doi:10.1088/1681-7575/aa4e59
	[arXiv:1607.03165 [physics.atom-ph]].
	%%CITATION = doi:10.1088/1681-7575/aa4e59;%%
	%5 citations counted in INSPIRE as of 26 Jun 2017

	 
	 	\bibitem{Mohr}
	P. J. Mohr and B. N. Taylor, Phys.\ Rev.\ Lett. {\bf 72} (2000) 351.
	
	\bibitem{Ericson}
	T. E. O. Ericson and M. Rosa-Clot, Nucl. Phys. {\bf A405} (1983) 497.
	
		    %\cite{Abbott:2000ak}
  \bibitem{Abbott:2000ak} 
  D.~Abbott \emph{et al}. [JLAB t20 Collaboration],
  %``Phenomenology of the deuteron electromagnetic form-factors,''
  Eur.\ Phys.\ J.\ A {\bf 7}, 421 (2000)
  doi:10.1007/PL00013629
  [nucl-ex/0002003].
  %%CITATION = doi:10.1007/PL00013629;%%
  %90 citations counted in INSPIRE as of 26 Jun 2017
  
%  %\cite{Bernauer:2016ziz}
%  \bibitem{Bernauer:2016ziz} 
%  J.~C.~Bernauer and M.~O.~Distler,
%  %``Avoiding common pitfalls and misconceptions in extractions of the proton radius,''
%  arXiv:1606.02159 [nucl-th].
%  %%CITATION = ARXIV:1606.02159;%%
%  %7 citations counted in INSPIRE as of 11 Sep 2018
%  
%  %\cite{Sick:2018fzn}
%  \bibitem{Sick:2018fzn} 
%  I.~Sick,
%  %``Proton charge radius from electron scattering,''
%  Atoms {\bf 6}, no. 1, 2 (2018)
%  doi:10.3390/atoms6010002
%  [arXiv:1801.01746 [nucl-ex]].
%  %%CITATION = doi:10.3390/atoms6010002;%%
%  %3 citations counted in INSPIRE as of 11 Sep 2018
  
  \bibitem{Minuit}
  F. James. Cern Program Library, {\bf D506} (1994). Minuit manual. \url{https://root.cern.ch/download/minuit.pdf}.
  
  %\cite{Yan:2018bez}
  \bibitem{Yan:2018bez} 
  X.~Yan {\it et al.},
  %``Robust extraction of proton charge radius from electron-proton scattering data,''
  Phys.\ Rev.\ C {\bf 98}, no. 2, 025204 (2018)
  doi:10.1103/PhysRevC.98.025204
  [arXiv:1803.01629 [nucl-ex]].
  %%CITATION = doi:10.1103/PhysRevC.98.025204;%%
  %3 citations counted in INSPIRE as of 11 Sep 2018
  
  %\cite{Jones:1999rz}
\bibitem{Jones:1999rz} 
  M.~K.~Jones {\it et al.} [Jefferson Lab Hall A Collaboration],
  %``G(E(p)) / G(M(p)) ratio by polarization transfer in polarized e p ---> e polarized p,''
  Phys.\ Rev.\ Lett.\  {\bf 84}, 1398 (2000)
  doi:10.1103/PhysRevLett.84.1398
  [nucl-ex/9910005].
  %%CITATION = doi:10.1103/PhysRevLett.84.1398;%%
  %853 citations counted in INSPIRE as of 11 Oct 2018

%\cite{Gayou:2001qd}
\bibitem{Gayou:2001qd} 
  O.~Gayou {\it et al.} [Jefferson Lab Hall A Collaboration],
  %``Measurement of G(Ep) / G(Mp) in polarized-e p ---> e polarized-p to Q**2 = 5.6-GeV**2,''
  Phys.\ Rev.\ Lett.\  {\bf 88}, 092301 (2002)
  doi:10.1103/PhysRevLett.88.092301
  [nucl-ex/0111010].
  %%CITATION = doi:10.1103/PhysRevLett.88.092301;%%
  %777 citations counted in INSPIRE as of 11 Oct 2018
  
  %\cite{Puckett:2010ac}
\bibitem{Puckett:2010ac} 
  A.~J.~R.~Puckett {\it et al.},
  %``Recoil Polarization Measurements of the Proton Electromagnetic Form Factor Ratio to Q^2 = 8.5 GeV^2,''
  Phys.\ Rev.\ Lett.\  {\bf 104}, 242301 (2010)
  doi:10.1103/PhysRevLett.104.242301
  [arXiv:1005.3419 [nucl-ex]].
  %%CITATION = doi:10.1103/PhysRevLett.104.242301;%%
  %243 citations counted in INSPIRE as of 11 Oct 2018
  
  %\cite{Punjabi:2005wq}
\bibitem{Punjabi:2005wq} 
  V.~Punjabi {\it et al.},
  %``Proton elastic form-factor ratios to Q**2 = 3.5-GeV**2 by polarization transfer,''
  Phys.\ Rev.\ C {\bf 71}, 055202 (2005)
  Erratum: [Phys.\ Rev.\ C {\bf 71}, 069902 (2005)]
  doi:10.1103/PhysRevC.71.055202, 10.1103/PhysRevC.71.069902
  [nucl-ex/0501018].
  %%CITATION = doi:10.1103/PhysRevC.71.055202, 10.1103/PhysRevC.71.069902;%%
  %416 citations counted in INSPIRE as of 11 Oct 2018

%\cite{Puckett:2011xg}
\bibitem{Puckett:2011xg} 
  A.~J.~R.~Puckett {\it et al.},
  %``Final Analysis of Proton Form Factor Ratio Data at $\mathbf{Q^2 = 4.0}$, 4.8 and 5.6 GeV$\mathbf{^2}$,''
  Phys.\ Rev.\ C {\bf 85}, 045203 (2012)
  doi:10.1103/PhysRevC.85.045203
  [arXiv:1102.5737 [nucl-ex]].
  %%CITATION = doi:10.1103/PhysRevC.85.045203;%%
  %128 citations counted in INSPIRE as of 11 Oct 2018
  
  %\cite{Ron:2011rd}
\bibitem{Ron:2011rd} 
  G.~Ron {\it et al.} [Jefferson Lab Hall A Collaboration],
  %``Low $Q^2$ measurements of the proton form factor ratio $mu_p G_E / G_M$,''
  Phys.\ Rev.\ C {\bf 84}, 055204 (2011)
  doi:10.1103/PhysRevC.84.055204
  [arXiv:1103.5784 [nucl-ex]].
  %%CITATION = doi:10.1103/PhysRevC.84.055204;%%
  %89 citations counted in INSPIRE as of 11 Oct 2018

\end{thebibliography}
\end{document}